\documentclass{aa}
\usepackage{natbib}
\usepackage{amssymb}
\usepackage{amsmath}
\usepackage{graphicx}

\def\aap{A\& A}

\def\araa{AnnRevA\& A}
\def\mnras{MNRAS}

\def\apj{ApJ}

\def\apjl{ApJL}
\def\apjs{ApJS}

\def\www{\ensuremath{{\cal W}}}
\def\sss{\ensuremath{{\cal S}}}
\def\wface{\ensuremath{{\cal W}_1}}
\def\wedge{\ensuremath{{\cal W}_2}}
\def\sface{\ensuremath{{\cal S}_1}}
\def\sedge{\ensuremath{{\cal S}_2}}
\def\isotr{\ensuremath{{\cal I}}}
\def\comma{\,,}
\def\fullstop{\,.}
\begin{document}
\title{Clumpy tori around active galactic nuclei}
\titlerunning{Clumpy tori around AGN}
\authorrunning{Dullemond \& van Bemmel}
\author{C.P.~Dullemond \& I.M.~van Bemmel\thanks{ESA 
Fellow, Space Telescope Division of ESA}}
\institute{Max Planck Institut f\"ur Astrophysik, P.O.~Box 1317, D--85741 
Garching, Germany; e--mail: dullemon@mpia-hd.mpg.de\\
Space Telescope Science Institute, 3700 San Martin Drive, Baltimore, MD 21218, USA}
\date{DRAFT, \today}

\abstract{We discuss the question whether the matter in dusty tori around
active galactic nuclei has a smooth or a clumpy structure.  Nenkova, Ivezic
\& Elitzur (\citeyear{nenkovaivez:2002}) have argued that the lack
of emission feature in the SEDs of type 1 AGN galaxies combined with a clear
absorption feature in type 2 AGN can be explained if the circumnuclear dust
is distributed in discrete clumps. Our aim is to verify this. We use
multi-dimensional radiative transfer models of smooth and clumpy tori, and
compare the SEDs of equivalent smooth and clumpy models. We find
that the 10 $\mu$m emission feature of the clumpy models, when seen almost
face-on, is not appreciably reduced compared to the equivalent smooth
models. Some of the clumpy models have a weak or even absent 10 $\mu$m
feature, but so do some of the smooth models. On the whole the SEDs of
clumpy and smooth tori are similar, but some details are different. The
absorption feature seen at edge-on inclinations appears to be less deep in
the clumpy models than in the smooth models, and the average flux in the
near-infrared regime is stronger in the clumpy models. Moreover, at these
inclinations the clumpy models have a slightly wider SED. Whether these
differences are unique enough to be used as a diagnostic for clumpiness of
AGN tori is not yet clear.}

\maketitle


\begin{keywords}
\end{keywords}

\section{Introduction}
According to the unification principle of active galactic nuclei (AGN), the
intrinsic difference between a Seyfert 1 and a Seyfert 2 galaxy (for the
radio-quiet AGN) or between a narrow-line radio galaxy and a radio
quasar (for radio-loud AGN) is merely a question of orientation (Antonucci
\& Miller \citeyear{antonmiller:1985}; Barthel \citeyear{barthel:1989}; see
review by Antonucci \citeyear{antonucci:1993}). The leading hypothesis is
that the central engine, an accreting super-massive black hole, is
surrounded by a geometrically and optically thick torus of dust and gas with
an equatorial visual optical depth much larger than unity. When viewed
face-on, the source would look like a type~1 active galaxy, e.g.~a Seyfert 1
or a quasar, and when viewed edge-on, it would have the characteristics of a
type~2 active galaxy, e.g.~a Seyfert 2 or a narrow-line radio
galaxy. From this unification model it follows that the dust in
the circumnuclear torus emits strongly in the infrared, due to the
irradiation by the central source. Indeed, the predicted infrared emission
from multi-dimensional continuum radiative transfer models for such tori is
reasonably well in agreement with the observed infrared radiation from such
active nuclei (Pier \& Krolik \citeyear{pierkrolik:1993}; Efsthathiou \&
Rowan-Robinson \citeyear{efstathiourowan:1995}; Granato, Danese \&
Francheschini \citeyear{granatodanese:1997}; van Bemmel \& Dullemond
\citeyear{vanbemmeldullemond:2003}, henceforth vBD03). Moreover, various
other differences between Seyfert 1 / Seyfert 2 and between quasar / radio
galaxy can be explained in terms of such a torus model. For instance, the
measured polarized (i.e.~reflected) nuclear emission in type~2 sources
proves that an active nucleus is present even though no direct emission is
observed (Antonucci \& Miller \citeyear{antonmiller:1985}; Pier et
al.~\citeyear{pieranton:1994}).

In spite of the success of the obscuring torus model, there are a
number of unsolved problems with this scenario. The most troubling
problem originates from the fact that, in order to have a
hydrostatically supported geometrically thick torus around a
supermassive black hole, the temperature of the torus must be of the
order of 10$^6$K or more. Dust in such a hot torus would not survive
long, yet dust signatures are observed to be present, such as a
pronounced mid- to far-infrared thermal bump and 10 $\mu$m Si-O
stretching band of silicate in absorption in type~2 sources. Various
solutions have been proposed in the past. Pier \& Krolik
(\citeyear{pierkrolikradpress:1992}, henceforth PK92) suggested that
radiation pressure within the torus may be enough to support
it. Dopita et al.~(\citeyear{dopitaheisler:1998}), on the other hand,
put forward the scenario that the torus is a slowly rotating
free-falling ``envelope'', that circularizes at the centrifugal radius
where it feeds the accretion disk around the black hole.  Another
scenario, first suggested by Krolik \& Begelman
(\citeyear{krolikbegel:1988}), is that the torus in fact consist of a
large number of optically thick clumps orbiting around the central
engine and experiencing regular collisions with other clumps. More
recently it was shown that a nuclear starburst could provide enough
energy input into the torus via supernovae, that the torus can keep up
its scale height and has a ``sponge'' like structure (Wada \& Norman
\citeyear{wadanorman:2002}). All of these scenarios have their
strengths and problems, and the issue is still subject of debate
(e.g.~Vollmer, Beckert \& Duschl \citeyear{vollmerbeckdusch:2004}).

In spite of lack of detailed knowledge about the structure of the
torus, several studies have tried to describe its emission properties,
using radiative transfer modeling of smooth tori (PK02; Efsthathiou 
\& Rowan-Robinson \citeyear{efstathiourowan:1995}, henceforth ERR95; 
Granato \& Danese \citeyear{granatodanese:1994}, henceforth GD94; 
vBD03). They encountered two major problems in matching the torus
emission models to the observations.  First, many of the model 
spectral energy distributions (SEDs)
were too narrow to fit the observed broad mid- and
far-infrared SEDs in active galaxies. This can either be related to
the presence of alternative infrared emission mechanisms, but it has
also been shown that the adopted radius of the torus affects the width
of the resulting SED (vBD03).

A second, and still largely unsolved, issue is that the 10 $\mu$m
silicate feature is often observed in absorption in type~2 sources,
but has never been observed in emission in either type~1 or 2 sources.
Radiative transfer models of smooth tori tend to predict a clearly
measurable 10 $\mu$m feature in emission for type~1 sources (PK92,
GD94, ERR95).  However, Laor \& Draine (\citeyear{laordraine:1993}) and 
vBD03 have shown that with larger grains
dominating the grain-size distribution the 10 $\mu$m feature is absent
in type~1's.  Recently, Nenkova, Ivezic \& Elitzur
(\citeyear{nenkovaivez:2002}, henceforth NIE02) proposed a different
explanation: they suggest that clumpy tori -- and only clumpy tori --
naturally have these desired properties. Their claim is based on a
model for a single clump irradiated by the central engine and by
neighboring clumps. A statistical generalization of this single-clump
model to a clumpy torus is made, and the SED
computed. The clump optical depth is taken as a global
parameter. They find that if their clump optical depth exceeds 60,
and the typical distance between clumps increases proportionally to
radius, then the behaviour of their clumpy model is in better
agreement with the observations than smooth torus models when it comes
to the 10 $\mu$m feature. Relatively few clumps (typically $\sim 5$)
are needed in the line of sight. In addition to this, they find that
the SED of such a configuration is relatively wide, in accordance with
observations.

While the properties of the clumpy torus model of NIE02 are
attractive, their model is highly approximative. First of all, their
single-clump model was computed using a 1-D radiative transfer code,
even though the main source of irradiation of the clumps near the dust
evaporation radius is clearly one-sided and requires at least a 2-D
axisymmetric approach.  Secondly, their statistical approach to the
generalization from one clump to an ensemble of clumps may be correct,
but remains unproven.

In this paper we take a first step toward a more self-consistent model 
and we will test the claim by NIE02 that infrared observations of active 
galaxies point to a clumpy torus. In order to do so, we
model the clumpy torus as a whole, using a multi-dimensional Monte-Carlo
radiative transfer program called {\tt RADMC}. Since {\tt RADMC} can only
handle axisymmetric problems (i.e.~2-D problems in $R$ and $\Theta$), our
``clumps'' are in reality rings around the polar axis. While this setup does
not constitute a realistic 3-D clumpy torus, it does have many of the
characteristics of such a torus: clumps can cool by radiating in all
directions, radiation can move freely between clumps  and there are 
high density constrasts. We therefore believe that this is a good first 
step toward an understanding of the properties of clumpy tori.

Our goal is to make a direct comparison between smooth models and
clumpy models with the same global physical parameters. If clumpiness
has a profound influence on the SED of a torus, this comparison should
yield distinct differences between the 10 $\mu$m feature and overall
width of smooth and clumpy torus models. The distribution of the
clumps is random, but on average the distribution of matter of the
clumpy torus is the same as in the smooth torus. Following NIE02 we
assume that all clumps have the same optical depth.

\section{Model setup}
We solve the problem of continuum radiative transfer through a dust density
distribution around an active nucleus of luminosity
$L_{\mathrm{agn}}=1\times 10^{11}L_{\odot}$. The spectral shape of the
nuclear emission is taken to be that used in the models of GD94, 
but the precise spectral shape does not
have a major effect on the results of model.  The distribution of dusty
matter around the nucleus is modeled on a computational grid based on
spherical (polar) coordinates $R$, $\Theta$ and $\Phi$. Since our radiative
transfer program {\tt RADMC} can only handle axially symmetric density
distributions, the model setup depends only on $R$ and $\Theta$. In addition
to this, we assume mirror symmetry in the equatorial plane located at
$\Theta=\pi/2$, i.e.~we only model the domain $0\le \Theta \le \pi/2$. The
radial grid is logarithmically spaced, i.e.~it has a constant $\Delta
R/R$. Such a grid ensures proper spatial resolution over a wide range of
radii, so that a torus with large ratio of outer over inner radius can be
modeled without resolution problems. The $\Theta$ grid is linearly spaced.
In order to properly resolve the clumps we need a high spatial resolution of
our grid. We have 356 $R$-grid points from the inner to the outer radius and
we have 120 $\Theta$-grid points from pole to equatorial plane.

Our global torus setup is kept very simple. The density
$\rho_{\mathrm{s}}(R,\Theta)$ for the smooth torus setup is a powerlaw
function of $R$, and is constant with $\Theta$ within a certain domain:
\begin{equation}
\rho_{\mathrm{s}}(R,\Theta) = \left\{\begin{matrix}
\rho_{\mathrm{s}0}\,(R/pc)^p & \mathrm{for} & \mathrm{abs}(\pi/2-\Theta)\le \Delta \\
0                & \mathrm{for} & \mathrm{abs}(\pi/2-\Theta)> \Delta \\
\end{matrix}\right.
\fullstop
\end{equation}
The quantity $\Delta$ is the geometric thickness of the torus. This
parameter determines the overall luminosity of the infrared emission of the
torus, but it has only weak effect on the shape of the SED. In this paper we
shall take it fixed at $\Delta=\pi/4$, corresponding to an opening
angle of 45 degrees. As inner- and outer radius we take, rather
arbitrarily, 0.3 and 10 parsec respectively.  The effect of varying inner
and outer radius is described in detail in vBD03. Choosing these values
differently will affect the width of the SED, but not so much the
10 $\mu$m feature. After optimizing the other model parameters, the inner
and outer radius can be adjusted to tune the width of the SED and $\Delta$
to tune to luminosity, in order to match observations.  Many of the
additional effects of the geometry of the torus have been described already
by vBD03 and by others (e.g.~ERR95; GD04), so we do not need to repeat all
of them here.

We present four smooth torus and sixteen clumpy models in this paper. For
each smooth model there are four clumpy models with the same global physical
parameters. Two of the clumpy models have 40 clumps, and two have 20
clumps. The only difference between the pairs of clumpy models with
identical number of clumps are the random positions of the clumps in
$R,\Theta$. The smooth models will be used as benchmarks against which the
clumpy models can be compared.  In this paper we will focus on the effect of
the radial powerlaw $p$ of the density distribution ($\rho\propto R^p$), and
study the effect of thermal decoupling between silicate and graphite
grains. 

\begin{table*}
\centerline{
\begin{tabular}{c|cccc|ccccc}
Model No.& $p$& GSD    & TD      & scat    & \wface  & \wedge  &
\sface   & \sedge  &  \isotr   \\
\hline
S1     &   0  & $0.25$ &         &         & 1.17 & 0.44  &  0.314 & -2.894 & 0.110 \\
S2     &   -1 & $0.25$ &         &         & 1.13 & 0.54  & -0.016 & -1.467 & 0.068 \\
S3     &   0  & MRN    & $\surd$ & $\surd$ & 1.26 & 0.41  &  0.137 & -2.025 & 0.090 \\
S4     &   -1 & MRN    & $\surd$ & $\surd$ & 1.13 & 0.51  & -0.083 & -1.906 & 0.060 \\
\end{tabular}}
\caption{\label{tab-smooth-models}Overview of parameters and results of the
smooth models. Columns from left to right: model number, powerlaw index $p$
for $\rho(R)\propto R^p$, grain size distribution (either $0.25\mu$m or MRN
distribution), temperature decoupling between silicate and carbon grains,
inclusion of scattering opacity, the resulting SED widths \wface{} (for
$i=20^{\mathrm{o}}$) and \wedge{} (for $i=90^{\mathrm{o}}$), the resulting 10
$\mu$m feature strengths \sface{} (for $i=20^{\mathrm{o}}$) and \sedge{} (for
$i=90^{\mathrm{o}}$) and the resulting anisotropy parameter \isotr{}.  See
text for definition of \wface{}, \wedge{}, \sface{}, \sedge{} and \isotr{}.}
\end{table*}

\subsection{Smooth torus description}
An overview of the smooth models is given in Table~\ref{tab-smooth-models}.
All models assume a 50\% carbon and 50\% graphite mixture for the dust.  For
the silicate opacity we use the optical constants of Laor \& Draine
(\citeyear{laordraine:1993}). The optical constants for amorphous carbon
were taken from Preibisch et al.~(\citeyear{prebishoss:1993}). All models
have a total dust mass in the torus of $2\times 10^6$\,M$_{\odot}$. We vary
$p$ between 0 and --1. We also vary the dust properties: on the one hand we
use thermally coupled 0.25 $\mu$m sized silicate and carbon grains without
scattering (the scattering opacity taken to be zero); on the other
hand we use a distribution of Galactic (MRN) dust between 0.005 and 1 $\mu$m
{\em with} a treatment of scattering (albeit in isotropic approximation). A
standard MRN distribution only extends up to 0.25 $\mu$m, but we chose to
extend it to 1 $\mu$m to maximize any possible effects of such a
distribution compared to the single-grain-size models. It should be noted
that we have only introduced a thermal decoupling between the carbon and
silicate grains, but not between the different sizes of the MRN
distribution, since the latter would be very computationally expensive, in
particular for the high spatial grid resolution required for the clumpy
models described below. We do not expect this to have much
effect, since grains up to 1 $\mu$m size all have the same 10 $\mu$m
feature shape. But future modeling will have to verify this.

\subsection{Clumpy torus description}
An overview of the clumpy models is given in Table~\ref{tab-clumpy-models}.
For the clumpy torus models we start from the smooth torus models
S1$\cdots$S4, and contract the matter into discrete annular clumps randomly
positioned on the $R$,$\Theta$ computational grid. The random positions are
distributed such that on average the density is the same as that of the
equivalent smooth torus. Following NIE02 we take the optical depth of the
clumps to be a global parameter of the model. Another global parameter is
the relative size of the clumps compared to $R$, i.e.~$\sigma\equiv
\mathrm{size}/R$.  This means that the size of the clumps scales with
distance from the black hole. For the models presented in this paper we take
this constant to be $\sigma=0.025$. Ideally we would wish to model smaller
clumps (and more of them), but technical limitations of the resolution of
our computational domain also limit the minimum size of our clumps, since it
is important that all clumps are well resolved by the grid. 

We present two pairs of clumpy models to match each smooth torus
model, the clumpy models are numbered accordingly, i.e. C1 equals S1,
etc. Between the pairs of clumpy models, only the number of clumps is
varied, which we denote with suffices a and b, the a-series always having 40
clumps and the b-series having 20. Within the pairs we vary the random
distribution of the clumps, allowing us to also study the effect of
randomness on the resulting SED. This is denoted with the number
following suffix a or b.

The matter within each individual clumps is distributed as follows:
\begin{equation}
\rho_{\mathrm{c}i}(R,\Theta) = \rho_{\mathrm{c0}i} 
\exp\left(-\frac{(R/R_i-1)^2}{\sigma^2}-\frac{(\Theta-\Theta_i)^2}{\sigma^2}\right)
\comma
\end{equation}
where $i$ stands for clump number $i$.

\subsection{SED generation}
Once the clumps are put onto the computational domain, and the density
distribution $\rho(R,\Theta)$ is set up, the continuum radiative transfer
problem is solved using a Monte-Carlo program called {\tt RADMC} (Dullemond
\& Dominik \citeyear{duldomdisk:2004}), which uses an improved version of
the original algorithm of Bjorkman \& Wood
(\citeyear{bjorkmanwood:2001}). This program solves the transport of
continuum radiation and the local thermal equilibrium of the dust grains, thus
obtaining the temperature of the dust everywhere on the grid. The SED at
inclinations of 20,70, and 90 degrees are then computed using a ray tracing
code. We choose $i=20^{\mathrm{o}}$ to be the representative
inclination for type 1 (face-on) AGN since with our sharp inner edge setup
a perfect $i=0^{\mathrm{o}}$ inclination would produce an artificially
low near-infrared flux because one would look perfectly along the inner
edge. The representative inclination for type 2 (edge-on) AGN is chosen
to be $i=90^{\mathrm{o}}$.

\section{Results and analysis}
In this section we present the results of the model calculations. We present
figures of the SEDs at different inclination angles. We analyze our results
using quantitative numbers for the width \www{} of the SED 
(\wface{} for face-on and \wedge{} for edge-on), the strength \sss{} of the
10 $\mu$m feature (\sface{} for face-on and \sedge{} for edge-on)
and the anisotropy parameter \isotr{}. Following PK92 and GD94 we define the
width \www{} (i.e.~\wface{} and \wedge{}) of the SED as the 10-log
of the frequency range in which the spectrum is more than 1/3 of its peak
value. In $\nu F_\nu$; for a pure blackbody spectrum this value is
\www$=0.686$, but observations indicate that the SED of active galaxies
is generally much wider.

The feature strength \sss{} (i.e.~\sface{} and \sedge{}) is
defined as the e-log of the peak-over-continuum ratio of the feature for
face-on inclinations.  Here an e-log is used, allowing a direct comparison
to previous studies (GD94, Laor \& Draine 1993).  Following GD94 the
continuum is defined by a powerlaw connecting the fluxes at 6.8 and 13.9
$\mu$m.  A positive value of \sss{} means the 10 $\mu$m feature is in
emission, a negative value means absorption.

Finally we define the isotropy parameter \isotr{} as the {\em linear} ratio
of the total integrated infrared flux at $90^{\mathrm{o}}$ inclination over
the total integrated infrared flux at $20^{\mathrm{o}}$ inclination. This
implies that for larger values of \isotr{} there is more isotropy,
\isotr{}$=1$ indicating perfect isotropy.

\begin{table*}
\centerline{
\begin{tabular}{c|ccccc|c|ccccc}
Model No. & $p$ & $N_{\mathrm{clump}}$ &
GSD & TD & scat &
$\tau_{\mathrm{clump}}$ & \wface & \wedge & \sface & \sedge & \isotr \\
\hline
C1\_a1 & 0  & 40 & $0.25$ &         &         &   26 & 0.91 & 0.57 & 0.347  & -1.585 & 0.176 \\
C1\_a2 & 0  & 40 & $0.25$ &         &         &   26 & 0.91 & 0.63 & 0.377  & -0.486 & 0.301 \\
C1\_b1 & 0  & 20 & $0.25$ &         &         &   53 & 0.83 & 0.60 & 0.330  & -0.396 & 0.250 \\
C1\_b2 & 0  & 20 & $0.25$ &         &         &   51 & 0.77 & 0.66 & 0.433  &  0.039 & 0.294 \\
\hline                                                                                                                                                                                                                                                      
C2\_a1 & -1 & 40 & $0.25$ &         &         &   60 & 0.96 & 0.63 & 0.170  &  0.068 & 0.089 \\
C2\_a2 & -1 & 40 & $0.25$ &         &         &   60 & 1.05 & 0.63 & 0.117  & -0.880 & 0.098 \\
C2\_b1 & -1 & 20 & $0.25$ &         &         &  105 & 1.00 & 0.66 & 0.318  & -0.373 & 0.126 \\
C2\_b2 & -1 & 20 & $0.25$ &         &         &  115 & 1.17 & 0.92 & 0.150  & -0.464 & 0.157 \\
\hline                                                                                                                                                                                                                                                      
C3\_a1 & 0  & 40 & MRN    & $\surd$ & $\surd$ &   53 & 1.00 & 0.55 & 0.260  & -0.953 & 0.161 \\
C3\_a2 & 0  & 40 & MRN    & $\surd$ & $\surd$ &   53 & 0.92 & 0.63 & 0.263  & -0.381 & 0.280 \\
C3\_b1 & 0  & 20 & MRN    & $\surd$ & $\surd$ &  109 & 0.92 & 0.60 & 0.256  & -0.257 & 0.240 \\
C3\_b2 & 0  & 20 & MRN    & $\surd$ & $\surd$ &  104 & 0.83 & 0.66 & 0.320  & -0.025 & 0.289 \\
\hline                                                                                                                                                                                                                                                      
C4\_a1 & -1 & 40 & MRN    & $\surd$ & $\surd$ &  123 & 1.05 & 0.63 & 0.131  & -0.190 & 0.086 \\
C4\_a2 & -1 & 40 & MRN    & $\surd$ & $\surd$ &  123 & 1.05 & 0.75 & 0.059  & -0.703 & 0.085 \\
C4\_b1 & -1 & 20 & MRN    & $\surd$ & $\surd$ &  214 & 1.09 & 0.66 & 0.245  & -0.336 & 0.121 \\
C4\_b2 & -1 & 20 & MRN    & $\surd$ & $\surd$ &  235 & 1.17 & 0.71 & 0.091  & -0.360 & 0.138 \\
\end{tabular}}
\caption{\label{tab-clumpy-models} Overview of the model parameters and
results of the clumpy models. Columns from left to right: model number,
powerlaw index $p$ for $\rho(R)\propto R^p$, number of clumps, grain size
distribution (either $0.25\mu$m or MRN distribution), temperature decoupling
between silicate and carbon grains, inclusion of scattering opacity, optical
depth of the clumps, the resulting width \wface{} of the face-on
SED, \wedge{} for the edge-on SED, the resulting 10 $\mu$m feature strength
\sface{} for the face-on SED and \sedge{} for the edge-on SED and
the resulting anisotropy parameter \isotr{}. See text for definition of
\wface{}, \wedge{}, \sface{}, \sedge{} and \isotr{}.  The optical depth of
the clumps follows from the number of clumps, the distribution of clumps and
the total mass of the torus, and is therefore not an independent parameter
of the model, hence the separated column for the optical depth.}
\end{table*}

\subsection{Smooth models}
\begin{figure*}
\centerline{\includegraphics[width=16cm]{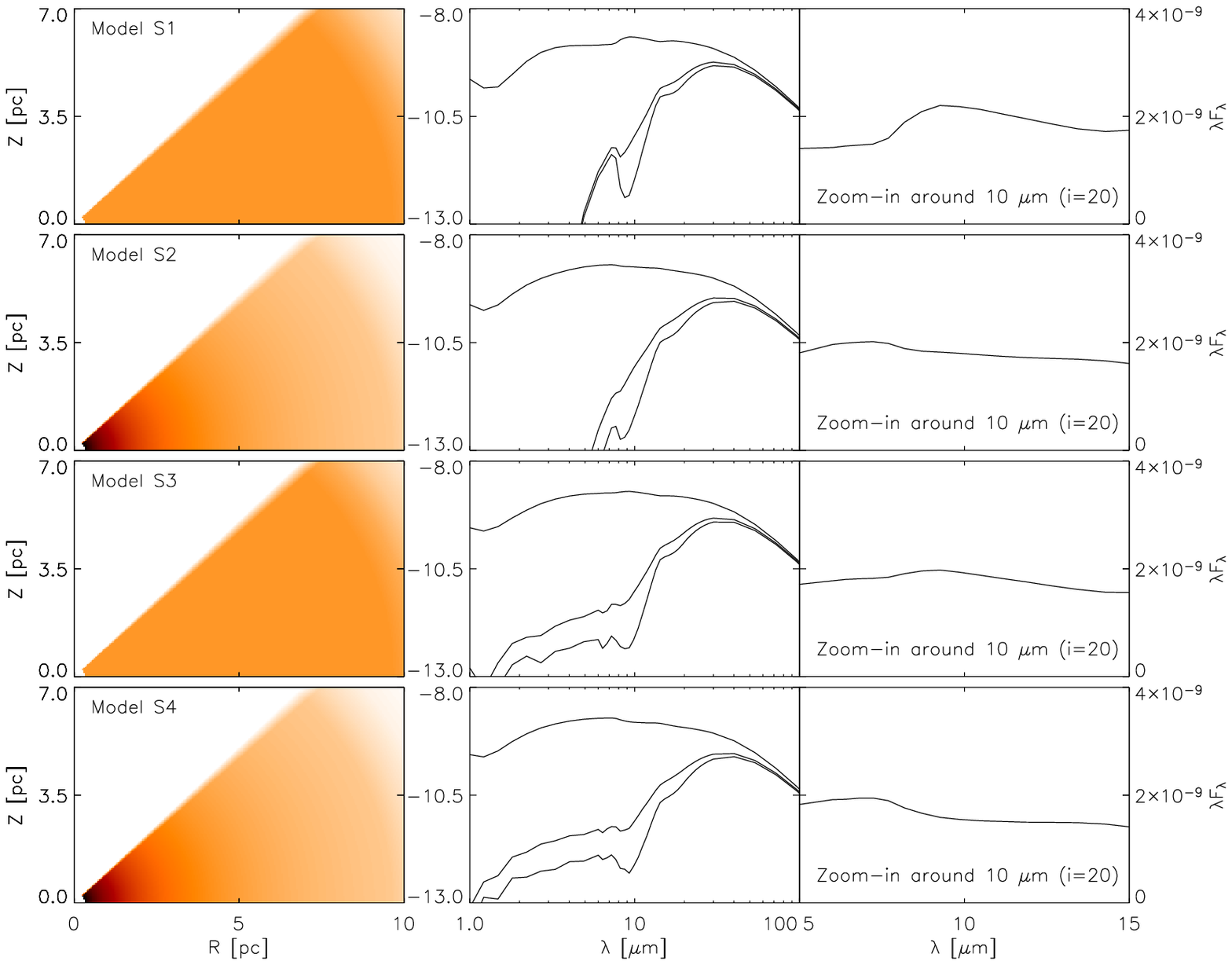}}
\caption{\label{fig-results-s}The results of the smooth torus models.  Each
row represents one model. The leftmost column represents the density
distribution. The middle column is the SED of the system at three
inclinations: 20$^o$, 70$^o$ and 90$^o$ (where 20$^{\mathrm{o}}$ is
effectively face-on, i.e.~corresponding to a perfect type~1 source and
90$^{\mathrm{o}}$ is edge-on, i.e.~corresponding to a perfect type~2
source.). The vertical axis represents $^{10}\log(\nu F_\nu)$ as seen at a
distance of $d=10^7$ pc. The SED includes the emission from the
torus as well as the emission from the illuminating star, though the latter
is only marginally visible in this figure. 
The right column is a linear plot of the face-on
(i=20$^o$) spectrum in the 10 $\mu$m regime, again $\nu F_{\nu}$ at $d=10^7$
pc.}
\end{figure*}

The results of the smooth torus models S1$\cdots$S4 are shown in
Fig.~\ref{fig-results-s}. We show the density distribution on the left, 
the total SED at different inclination angles in the middle, and on the 
right a zoom-in on the 10 $\mu$m region for face-on inclination.

For model S1 we have taken a constant density throughout the torus, because
according to GD94 such models fit better the observed
SEDs of AGN. We find that the SED has a width of \wface{}$=1.17$
(see table \ref{tab-smooth-models}).  The model produces a quite strong 10
micron emission feature for face-on tori (\sface{}$=0.314$), which is
inconsistent with observations. This strong emission feature is not entirely
surprising, because the torus is not very optically thick.  At $R=1$pc the
10 $\mu$m vertical optical depth through the torus is 2.9, and just
shortward of the feature at 7 $\mu$m it is 1.0. In fact, the nuclear
radiation impinging on the inner rim of the torus only reaches its $\tau=1$
surface (at $\lambda\sim 1\mu$m) at a radius of about $R=0.42$ pc
which is about 1.5 times the radius of the inner rim. Therefore, the inner
regions of the torus are relatively optically thin, and expected to produce
a strong emission feature.

Model S2 is identical to S1 except for changing $p=-1$. Effectively, this
puts more mass in the innermost regions of the disk, making these regions
more optically thick. As a result of this, the 10 micron emission feature
virtually disappears (\sface{}$=-0.016$). The SED also becomes a
tiny bit narrower (\wface{}$=1.13$). These effects are similar to
what was already found by PK92: compact tori with very high optical depth
can have a very weak emission feature, but produce narrower SEDs (see also
discussions in GD94). In the present case our torus is still relatively
large and hence not really compact. The large outer radius ensures that
there is a large reservoir of cool dust, which is relatively unaffected by
the change in $p$.  This is why the width is not strongly affected when
varying $p$.  The 10 $\mu$m feature, on the other hand, comes from the warm
inner regions (i.e.~small radius), which are very optically thick,
hence the disappearance of the 10 $\mu$m feature.

It should be noted, however, that the disappearance of the 10
$\mu$m feature for highly optically thick inner regions of the torus is not
obvious. It depends strongly on the geometry of these inner regions.  In our
conical torus model the edges are very sharp and straight. There is only one
surface of the torus that is directly irradiated, which is the inner rim.
When this irradiated hot inner rim is seen under an inclination of only
$i=20^{\mathrm{o}}$, as in the figures, then the line of sight toward a
surface element of the rim has a high inclination ($i\sim 70^{\mathrm{o}}$)
with respect to the normal vector on the rims surface, i.e.~almost parallel
to the surface. This weakens the emission feature. Moreover, the emission
from the near side of the rim is partly re-absorbed again by the rim
material itself. Additionally, a significant fraction of the 10 $\mu$m flux
comes from larger radii than the inner rim radius. That emission has (if
anything) an absorption feature. The end effect is that if our torus is
optically thick near the inner rim, the emission feature is very weak. The
total flux from the torus (including the regions at larger radii) may even
have a slight {\em absorption} feature, even for $i=20^{\mathrm{o}}$, as can
be seen in model S2 with \sface{}$=-0.016$.

Models S3 and S4 are like model S1 and S2 respectively, but now with
decoupling the temperatures for graphite and silicate grains, introducing an
MRN grain size distribution and including isotropic scattering. We
find that the 10 $\mu$m emission feature is significantly weakened
in S3 compared to model S1 (\sface{}$=0.137$ instead of
\sface{}$=0.314$), but still clearly present. The reason for the
weakening is that the graphite has a higher opacity at visual wavelengths
than silicate. Therefore the graphite becomes hotter than the silicate, and
produces more continuum emission in the 10 $\mu$m wavelength region. In
model S4 we see the silicate feature slightly in absorption
(\sface{}$=-0.083$) for $i=20^{\mathrm{o}}$ again.  The width of the
SED is virtually unaffected by changing to a proper dust grain size
distribution. If anything, S3 is somewhat wider than S1, but no difference
is found between S4 and S2. The same goes for the anisotropy: S3 is slightly
more anisotropic than S1, but S4 and S2 do not differ. 

\subsection{Clumpy models}
\begin{figure*}
\centerline{\includegraphics[width=16cm]{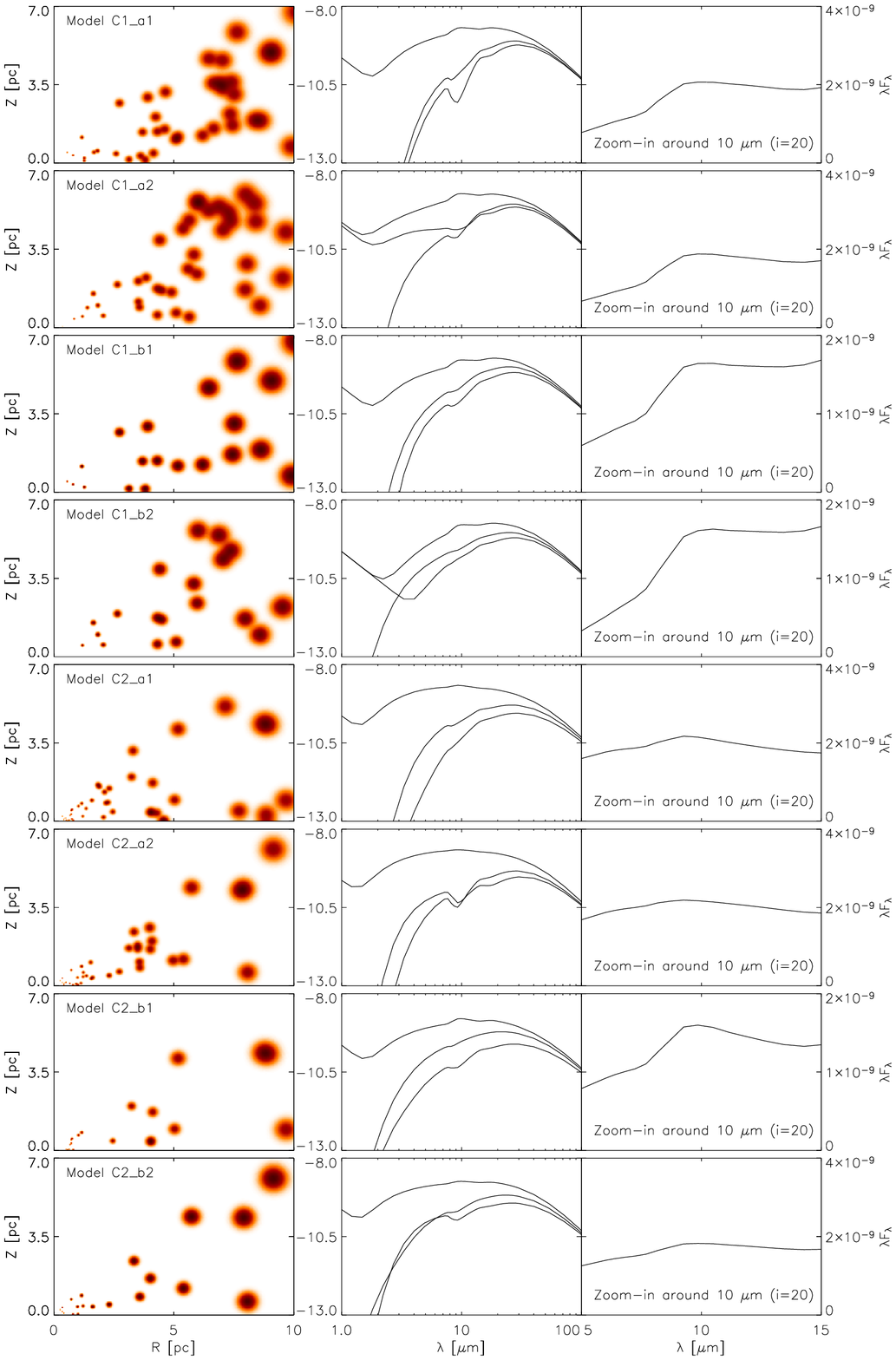}}
\caption{\label{fig-results-c12}The results of the clumpy torus models
C1 and C2. See Fig.~\ref{fig-results-s} for explanation. In 
contrast to Fig.~\ref{fig-results-s}, however, the left panel shows
$\rho\cdot R$ instead of $\rho$, so that all clumps have the same
grey depth.} 
\end{figure*}

\begin{figure*}
\centerline{\includegraphics[width=16cm]{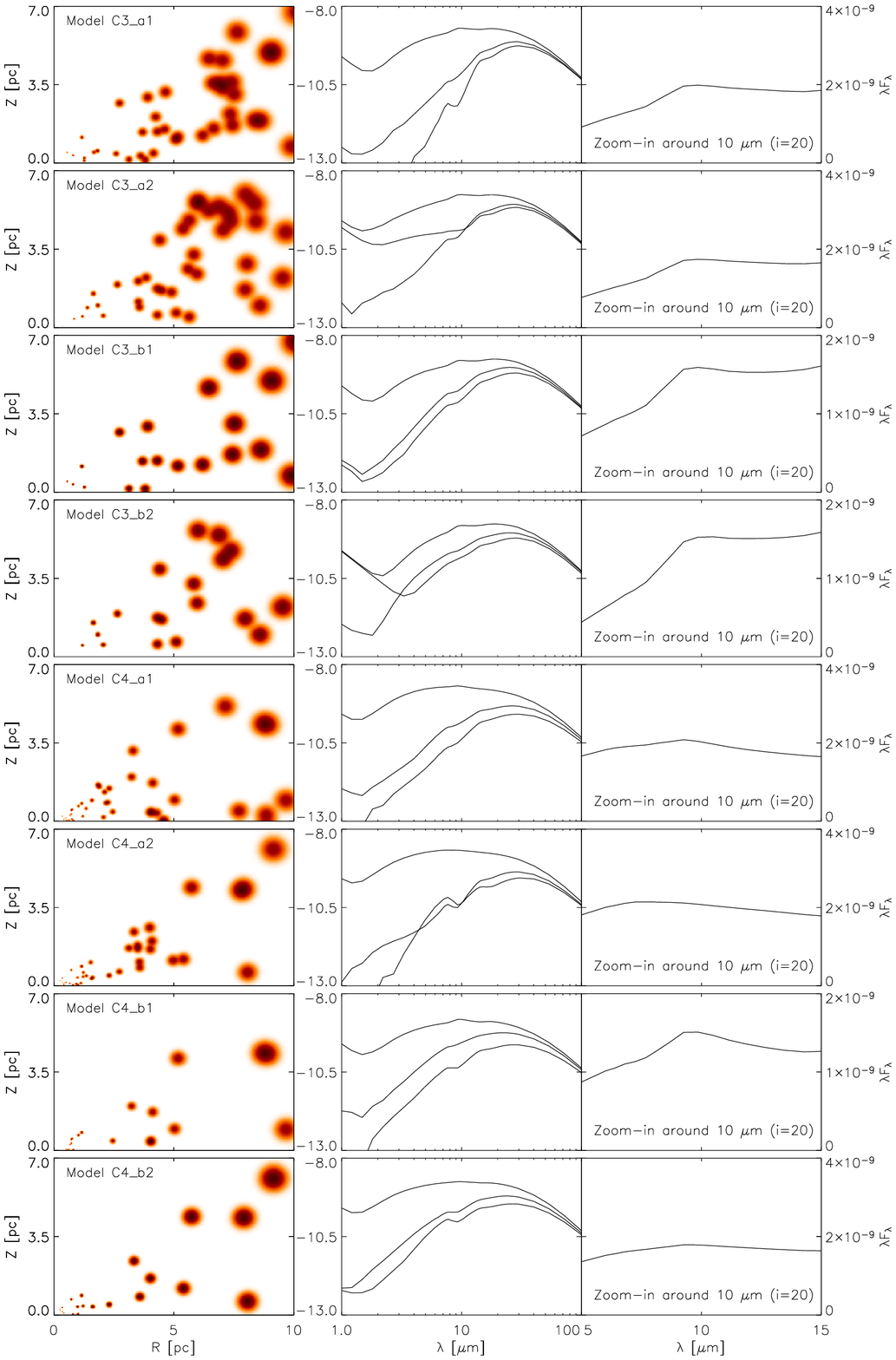}}
\caption{\label{fig-results-c34}As Fig.~\ref{fig-results-c12}, but 
for models C3 and C4.} 
\end{figure*}

The clumpy models C1$\cdots$C4 are the clumpy generalizations of models
S1$\cdots$S4. 
The results for the quantities \wface{},
\wedge{}, \sface{}, \sedge{} and \isotr{} are given in
Table~\ref{tab-clumpy-models}. The SEDs are shown in
Fig.~\ref{fig-results-c12} for models C1 and C2, and in
Fig.~\ref{fig-results-c34} for models C3 and C4. Again, on the left the
density distribution, middle the total SED at different inclination angles,
and on the right a zoom-in on the 10 $\mu$m region.

Figs.~\ref{fig-results-c12},\ref{fig-results-c34} show that with the
still relatively large size of the clumps ($\sigma=0.025$) and the
number of clumps used in these models, the distance between clumps is
not always very much larger than the size of the clumps
themselves. The NIE02 definition of a clumpy torus requires that the
clumps are very small compared to the mean free path between the
clumps. In this respect our model is not a true representation of
clumpy tori as defined by NIE02. However, as mentioned above,
technical limitations prevent us at present from modeling smaller and
more clumps. We use the a- and b-series to understand the effect of
the inter-clump distance, which is larger in the b-series.

The values of \wface{} do not seem to follow a major trend, except
perhaps that models C2 and C4 produce marginally broader SEDs than models C1
and C3. This is the opposite effect of what we observe in the smooth
models, where S2 and S4 generate narrower SEDs than S1 and S3. The isotropy
\isotr{} has a somewhat stronger trend with the $p=-1$ (`b') models being
more isotropic than the $p=0$ (`a') models, and the models C1 and C3 being
more isotropic than models C2 and C4.

The random location of the clumps can
have a reasonably strong effect on the values of \wface{}, \wedge{} and
\isotr{} and even more so on the values of \sface{} and \sedge{}. The reason
for these trends is more difficult to understand than for the smooth
models. 

\subsection{Comparison between smooth and clumpy models}
The face-on SED of models C1 are clearly less wide ($0.77 < \wface{} <
0.91$) than for model S1 (\wface{}$=1.17$), despite the same
global parameters.  The same is true for model C3 compared to
  S3. The difference in \wface{} between C2-S2 and C4-S4 is
  less pronounced, but still present. For edge-on inclinations the clumpy
  models have clearly wider SED than the smooth models with e.g.~model
  S1 having \wedge{}$=0.44$ and the C1 models having \wedge{} ranging
  between $0.57$ and $0.66$.
%

The clumpy models have a more isotropic emission, as can be seen
from the values of \isotr{}. For instance, the model S3 has 
\isotr{}$=0.09$
while the models C3 have values ranging between \isotr{}$=0.16$ and
\isotr{}$=0.29$.

%
The strength of the 10 $\mu$m emission feature for the $i=20^{\mathrm{o}}$
inclination does not seem to be decreased in the clumpy models. In
fact, on average the value of \sface{} is somewhat larger for the clumpy
models than for the smooth models. It is striking that while the S2 and S4
models have the feature slightly in absorption (\sface{}$=-0.016$ and
\sface{}$=-0.083$ respectively), the corresponding clumpy models
C2\_xx and C4\_xx have it consistently in emission: for the case C2\_b1 even
strongly in emission (\sface{}$=0.32$). So the clumpiness apparently
does little to suppress the emission feature. It {\em does},
however, appear to weaken the absorbtion feature for the high inclinations
(type 2 AGN). For example, the model S3 has \sedge{}$=-2.03$, which is a
strong absorption feature, while the models C3\_xx have values ranging
between $-0.95$ and $-0.03$.

\section{Discussion}

\subsection{Silicate feature strength in clumpy tori}
The results of our models indicate that clumpiness does not seem to
have the effect of suppressing the 10 $\mu$m silicate feature.  In fact, the
smooth models appear to do better in this respect than the clumpy models.
Some of our clumpy models do have a suppressed 10 $\mu$m feature
(e.g.~model C4\_a2 with \sface{}$=0.06$), but that seems to be
strongly influenced by the random location of the clumps (model C4\_a1 has
\sface{}$=0.13$). One reason why the C2 and C4 clumpy tori have on
average stronger 10 $\mu$m emission features in their face-on SEDs than the
smooth models S2 and S4 (which have no emission feature) is that the clumps
have a `fluffy' photosphere while the smooth models have sharp
edges. Emission features are best produced if a photosphere is irradiated
under a reasonably small angle, which is the case for a significant portion
of the fluffy photosphere of the clumps. Another reason is that in the
smooth models a part of the emission from the inner rim (in particular the
emission from the near side) is re-absorped by the torus, producing an
absorption feature there. For the clumpy tori this effect is weaker.



The weaker {\em absorption} feature for edge-on clumpy tori
compared to the edge-on smooth tori appears to be due to the fact that for
clumpy tori one can see at least partly between the clumps deeper into the
torus where the 10 $\mu$m emission is produced. Moreover, each of the clumps
in the outer regions of the torus, having a higher density than the
equivalent smooth torus, either blocks the light from the inner regions
entirely (if it is in the line of sight) or does not block any light (if it
is out of the line of sight). Only in few cases does one look just through
the fluffy photosphere of the clump, causing a 10 $\mu$m absorption
feature. In other words: the line-of-sight extinction caused by a collection
of clumps is more 'grey' than that of smoothly distributed matter.

It should be kept in mind that our results may depend on the
opacity we use. To allow a direct comparison with the NIE02 models, we have
chosen a Galactic grain size distribution and do not use larger grains than
1 $\mu$m. Grain sizes and distribution can have a profound effect on the
silicate feature, as shown by Laor \& Draine
(\citeyear{laordraine:1993}) and vBD03. There is convincing observational
evidence that the dust in active galaxies does not have Galactic properties
(Maiolino et al.~\citeyear{maiolinomarconi:2001}). Also, in radio galaxy
NGC\,4261 the near-infrared colours of the observed 300 pc scalee
disk cannot be modeled with standard Galactic dust (Martel et
al.~\citeyear{martelturnersparks:2000}). Therefore, it is quite
conceivable that the silicate feature is in reality weakened by opacity
effects. On the other hand, it is also important to ask the question whether
our simplification of thermally coupling the different grain sizes would
perhaps artificially suppress the feature.

\subsection{Width of the SEDs}\label{subsec-width}
As mentioned above, the width of the face-on SED \wface{} is
slightly smaller for the clumpy models than for the smooth models, in
particular for the $p=0$ models (S1,S3,C1,C3).  This effect can be explained
by the fact that for $p=0$ the typical number of clumps per $\delta \log R$
is not constant. For these models the clumps are typically more
concentrated to the outer regions of the torus. Since the number of clumps
is relatively moderate, the chance is then high that there will be few (if
any) clumps in the inner regions. The real inner radius of the torus is
determined by the location of the clumps. Therefore this deficiency of
clumps effectively shifts the inner radius of the clumpy torus
outward. If models would be made with $p<-1$, the same reasoning
would be applicable for the outer radius of the torus.  Since the width of
the face-on SED depends largely on the ratio of the outer to the
inner radius (see e.g.~GD94, vBD03), the effect of this shift will be to
narrow the SED. For the case of $p>-1$ the short-wavelength part of the SED
will be suppressed and for the case of $p<-1$ the long-wavelength part of
the SED will be suppressed.

On the other hand, for the edge-on SED one would expect that clumpy
tori have less absorption of the near-IR flux because emission from the
inner regions of the torus can travel between clumps toward the edge-on
observer. This effect is indeed seen in the models, which have larger
\wedge{} values for the clumpy models than for the smooth models.

\subsection{Isotropy of infrared emission}
Like a smooth torus, the bolometric infrared flux of 
a clumpy torus is stronger in the
polar direction than in equatorial directions. This is because in both cases
the emission from the hotter inner regions cannot be directly
observed for edge-on systems due to the obscuration by the cooler outer
regions. This anisotropy of the infrared emission is typical for all disk-
or torus-like configurations. Our models of clumpy tori show that their SEDs
are generally more isotropic than their smooth counterparts. Some models
show this stronger than others. In particular we find this effect to be very
strong for models C1 compared to model S1, while the effect is much less
pronounced for models C2 compared to model S2. The reason for the
increased isotropy of clumpy models is that radiation can freely move in
between clumps. In a way, the inter-clump distance acts as a new kind of
mean-free-path, and since there are not so many clumps in the model, the
new effective optical depth for the clumpy torus is therefore smaller
(even though the actual optical depth along individual lines-of-sight
may be still high). This effect is also reflected in the fact that the
edge-on mid-infrared flux for the clumpy models is significantly
higher than for the smooth models.

The increased isotropy in clumpy models C1 and C3 compared to C2 and C4 is
for a large part due to the fact that there were not so many clumps at small
radii, and therefore the inner radius was effectively moved outward (as
discussed in subsection \ref{subsec-width} above). Since the total mass of
the torus was kept constant (at $10^6\,M_{\odot}$) and the number of clumps
was kept the same ($N=$40 resp.~$N=$20), the optical depth of the clumps is
lower for the models C1 and C3 ($p=0$) than for the models C2 and C4
($p=-1$). A lower optical depth increases the isotropy of the torus (a
perfectly optically thin torus being perfectly isotropic). The generally
higher isotropy of the $N=$20 models versus the $N=$40 models (in spite of
the lower clump optical depth of the latter) is because for fewer clumps the
inter-clump distance is larger and the effective optical depth of the
torus is decreased.  

\subsection{A criterion for clumpiness?}
It is interesting to ask if, instead of the 10 $\mu$m silicate
emission feature, some of the other aspects of the infrared spectra of tori
could give clear indications for clumpiness. For example, for edge-on
inclinations (type 2 AGN) the width of the SED, the mid- over far-infrared
flux ratio and the depth of the absorption feature do show some trend with
clumpiness: the clumpy models having a slightly wider SED, higher
mid-infrared flux and shallower absorption feature. Unfortunately the
current study is not comprehensive enough to assure that these effects could
not be reproduced by other properties of the torus, such as the opacity
properties or the torus geometry. One has to scan a large parameter space of
smooth models to make sure that some property of the spectra of clumpy tori
is really unique to clumpy tori.  This is, however, beyond the scope of this
paper.

\subsection{The nature of clumps and clumpiness}
The models presented in this paper are meant to verify what is the effect of
clumpiness on the SED of torus models for active galaxies. Yet how
representative are our models for clumpy tori? Our clumps are not real 3-D
clumps due to our 2-D approximation, but we have argued that this should not
have a major effect on our conclusions. Aside from the 2-D issue, is our
description of the clumps realistic? Very little is known about the
structure of the dusty circumnuclear matter in active galactic
nuclei. Arguments for clumpiness have so far been rather indirect, but it
seems reasonable to assume that the circumnuclear matter is distributed in
an irregular and chaotic way rather than in a smooth and ordered manner.

The {\em kind} of clumpiness would depend much on the mechanism causing the
clumps. According to the model of Krolik \& Begelman
(\citeyear{krolikbegel:1988}) these clumps are individual dynamically
independent objects orbiting the black hole, and experiencing regular
semi-elastic collisions. These clumps must be very compact,
self-gravitating, and must be supported by strong interal magnetic fields to
provide sufficient elasticity upon collisions with other clumps. On the
other hand, a supersonically turbulent medium of the kind described by Wada
\& Norman (\citeyear{wadanorman:2002}) would produce filamentary (sponge)
structures rather than isolated clumps. These different structures of
clumpiness may have very different infrared emission properties. For
instance, a filametary medium is likely to have more matter in a marginally
optically thin state than a medium consisting of very compact clumps. Since
an emission feature comes from marginally optically thin regions, such an
emission feature is expected to be stronger for the filametary medium than
for the compact clumpy medium.

In the light of this, it is interesting to question what the effect is of
the ``fluffiness'' of the clumps in our simulation. In contrast to NIE02 we
assume our clumps to have a Gaussian density profile. If we would take
constant density clumps with a sharp edge, like NIE02, these clumps may have
less marginally optically thin material at their surface, perhaps
suppressing thereby the 10 $\mu$m emission feature where our models exhibit
this feature clearly in emission.  On the other hand, even for a perfectly
sharp edge of an optically thick clump, it is not guaranteed that the
feature vanishes because the surface of such a clump may be super-heated
with respect to the clump interior by the irradiation, yielding a hot
optically thin emission-feature-producing layer similar to what was
described for flared circumstellar disks by Chiang \& Goldreich
(\citeyear{chianggold:1997}). It is, unfortunately, not possible for us to
investigate this with our current models because this would require us to
increase the resolution of our computational grid by a large factor in order
to make sure to sample the photosphere of the clumps properly. This would be
prohibitively computationally expensive at present. We can therefore only
draw conclusions about clumpy media with Gaussian clumps.

Another issue related to this is the assumption, made by both NIE02 and
ourselves, that all the clumps have equal optical depth. According to Krolik
\& Begelman (\citeyear{krolikbegel:1988}) the clumps in a circumnuclear
torus get regularly tidally disrupted, forming smaller clumps which
subsequently merge to form bigger ones. In effect an equilibrium
distribution of clump sizes will result, with clumps of various sizes (and
optical depths) coexisting within the same torus. Some of these clumps may
easily be optically thin, or at least have low optical depth. Such clumps
may again provide a reservoir of marginally optically thin material which
could produce a 10 $\mu$m feature in emission. How strong this effect will
be depends on the equilibrium distribution function of clump sizes.

Finally it is important to mention that due to technical limitations we
could only model a rather limited number of clumps, each with a rather large
size. It cannot be excluded that some results may change if we would be able
to model problems with a much higher number of clumps, all of which being
much smaller than we have assumed in the models we presented here. It is
hard to estimate how big these effects are. In the limit of increasingly
many ever smaller clumps, while keeping the total mass of the torus and the
clump filling factor (the average number of clumps along the line of sight
toward the center) constant, the clump optical depth eventually drops below
unity. In this case the SED would become identical to that of the smooth
version. If one, on the other hand, keeps the optical depth of the clumps
constant while increasing the number of clumps, the average number of clumps
along the line of sight drops below unity, which would be against the whole
idea of obscuring circumnuclear tori. One would have to increase the total
mass of the torus to compensate for this. Since it poses technical problems
to model much smaller clumps than we have done in this paper, we cannot be
certain what the effect of such an increase of the number of clumps would
be. But by comparing the models with $N=40$ and $N=20$ (the a-series to the
b-series) we find very little differences, so we expect this to remain this
way for very high $N$.

\section{Conclusion}
We present the first global simulations of clumpy tori around AGN using a
axisymmetric, multi-dimensional radiative transfer model. From our
analysis and comparison between smooth and clumpy tori models we conclude
that the 10 $\mu$m feature can both be strengthened and weakened when
clumpiness is introduced. The width of the SED is largely determined by the
inner and outer radius and the main effect of clumpiness is to increase the
effective inner radius and/or decrease the effective inner/outer radius due
to statistical fluctuations in the positioning of the clumps. This results
in a slightly narrower face-on SED. The edge-on SED for
clumpy tori is slightly wider than for smooth tori, because radiation can
move freely between the clumps and emission from the inner regions of the
clumpy torus can more easily reach the observer even for edge-on
inclinations.  We find that the isotropy of the infrared emission is
significantly affected by clumpiness for similar reasons.
Unfortunately this is a quantity that cannot be directly observed in an
individual source. It requires studies of large samples of sources at
different inclinations with comparable physical properties.

We do confirm that for the particular parameters of clumpiness mentioned in
NIE02 the 10 $\mu$m feature can be rather weak, but this is even
more pronounced for a smooth torus with the same global parameters and
average density. A stronger effect is the depth of the 10
$\mu$m {\em absorption} feature for edge-on tori: for clumpy tori it is
clearly less deep than for smooth tori.

Although we use clumps with a slightly different structure than NIE02, and
although the clumps in our 2-D models are annuli around the symmetry axis
instead of real 3-D clumps, we believe that our models produce at least
qualitatively the correct results for clumpy media. We therefore
cast doubt on the idea that the properties of the 10 $\mu$m feature of type
1 and type 2 active galaxies point unequivocally to clumpy tori.  It should
be clear, though, that we do {\em not} claim that the circumnuclear matter
is smooth. We merely call for caution in interpreting the
properties of the SEDs in the context of clumpiness of the dusty tori in
active galaxies.

\begin{acknowledgements}
We wish to thank the referee, Ari Laor, for helpful comments.
CPD thanks Marc Schartmann for useful discussions.
\end{acknowledgements}

\end{document}